%Paper: hep-th/9301091
%From: Boguslaw Broda <PTBB@IBM.RZ.TU-CLAUSTHAL.DE>
%Date: Fri, 22 Jan 93 10:19:59 MET
%Date (revised): Sat, 20 Feb 93 16:28:13 MET

%""""""""""""""""""plaintex""""""""""""""""""""""B.Broda""""
% Chern-Simons approach to three-manifold invariants,
% by B.Broda (U.Clausthal and U.Lodz), 10 pages,
% plaintex (magstep 1), revised version (February 1993)
% preprint hep-th/9301091, January 1993
% supported by AvH and KBN
%
%""""""""definitions"""""""""macros"""""""""parameters""""""

\magnification \magstep1

\hoffset 1.5truecm
\hsize 16truecm \vsize 23.5truecm
\baselineskip 20pt \parskip5pt
\raggedbottom

\def\ml{{{\cal M}_L}}
\def\sthree{{{\cal S}^3}}
\def\tr{{\rm Tr}}
\def\half{{1\over2}}
\def\hol#1{{\rm Hol}_{K_{#1}}(A)}
\def\cd{{\cal D}}
\def\co{{\cal O}}
\def\wilson#1{W_{\mu_{#1}}^{K_{#1}}(A)}
\def\hg{\hat G}
\def\prodin{\prod_{i=1}^N}

\def\sigmami#1{\sum_{\mu_{#1}\in\hg}}
\def\partition{Z(\sthree,L;g_1,g_2,\dots,g_N)}
\def\avewilson#1#2{\left<W_{#1}^{#2}(A)\right>}
\def\sigmaad{\sum_{a=1}^d}
\def\tensorpr#1#2{t_{#1}^a\otimes t_{#2}^a}
\def\doublespace{\par\null\par}
\def\one{{\bf 1}}

\newcount\currentnumber
\def\currentno{\global\advance\currentnumber by 1
\the\currentnumber}

\def\heading#1{\doublespace\par\penalty-100
\noindent{\bf\currentno.~#1}\par\noindent}

\def\eqn#1{{\rm (\the\currentnumber.#1)}}

% skeins

\def\plusskein{
\setbox0=\hbox{
\setbox1=\hbox{\bigg/}
\setbox2=\hbox{$\backslash$}
\hbox to -\wd1{} \copy1
\kern-\wd1\raise.5\ht1\copy2
\lower\dp1\hbox{\raise\dp2\copy2}
}\copy0}

\def\minusskein{
\setbox0=\hbox{
\setbox1=\hbox{$\bigg\backslash$}
\setbox2=\hbox{/}
\hbox to -\wd1{} \copy1
\kern-\wd1\lower.5\ht1\copy2
\raise\dp1\hbox{\lower\dp2\copy2}
}\copy0}

\def\zeroskein{
\;\bigg\vert\;\bigg\vert\;}

\def\nullskein#1{
\;\bigg\vert #1\;}

%""""""""""""""""""text starts here"""""""""""""""""""""""""

\rightline{hep-th/9301091}
\vfill
\centerline{{\bf Chern-Simons approach to three-manifold
invariants}%
\footnote{$^\ast$}{Revised version (February 1993).}
}\smallskip
\centerline{BOGUS\L AW BRODA%
\footnote{$^\star$}{Humboldt fellow.}}
\smallskip
{\it
\centerline{Arnold Sommerfeld Institute for Mathematical Physics}
\centerline{Technical University of Clausthal, Leibnizstra\ss e 10}
\centerline{D-W--3392 Clausthal-Zellerfeld, Federal
Republic of Germany}
\smallskip
\centerline{\rm and}
\smallskip
\centerline{Department of Theoretical Physics, University
of \L\'od\'z}
\centerline{Pomorska 149/153, PL--90-236 \L\'od\'z, Poland\/%
            \footnote{$^\dagger$}{\rm Permanent address.}}
}
\vfill
%\eject
A new, formal, non-combinatorial approach to invariants of
three-dimensional manifolds of Reshetikhin, Turaev and
Witten in the framework of non-perturbative topological
quantum Chern-Simons theory, corresponding to an arbitrary
compact simple Lie group, is presented. A direct
implementation of surgery instructions in the context of
quantum field theory is proposed. An explicit form of the
specialization of the invariant to the group SU(2) is
derived, and shown to agree with the result of Lickorish.
\vfill
\noindent
{\bf AMS subject classifications (1991):} 57M25, 57N10,
57R65, 81T13.
\vfill
\centerline{JANUARY 1993}
\vfill\eject

\heading{Introduction}
Four years ago in his famous paper on quantum field theory
and the Jones polynomial [1], Edward Witten proposed a new
interesting topological invariant of {\it
three-dimensional\/} manifolds. An explicit construction
of the invariant, using quantum groups, appeared for the
first time in a paper of Reshetikhin and Turaev [2]. Other
papers presenting re-derivations of this result are more
geometrical by nature [3], and use the Temperley-Lieb
algebra, as suggested by Lickorish [4--6]. All the
approaches are combinatorial. Non-combinatorial
possibilities, very straightforward though mathematically
less rigorous, are offered by {\it topological} quantum
field theory [7].

Inspired by the papers [6,8], we aim to propose a new,
formal, non-combinatorial derivation of the three-manifold
invariants of the Reshetikhin-Turaev-Witten (RTW) type in
the framework of non-perturbative (topological) quantum
Chern-Simons (CS) gauge theory. The idea is extremely
simple, and in principle applies to an arbitrary compact
(semi-)simple group $G$ (not only to the SU(2) one). Our
invariant is essentially the partition function of CS
theory on the manifold $\ml$, defined via {\it surgery} on the
framed link $L$ in the three-dimensional sphere $\sthree$.
Actually, surgery instructions are implemented in the most
direct and literal way. The method of cutting and pasting
back, which has been successfully applied to
two-dimensional Yang-Mills theory [8], is explicitly used
in the standard field-theoretical fashion. Roughly
speaking, cutting corresponds to fixing, whereas pasting
back to identification and summing up the boundary conditions.

As a by-product of our analysis we consider the {\it satellite
formula}, and derive the {\it Kauffman bracket} polynomial
invariant of a trivial (with zero framing) unknot for an
arbitrary representation of SU(2).

\heading{General Formalism}
Our principal goal is to compute the partition function
$Z(\ml)$ of CS theory on the manifold $\ml$, defined via
(honest/integer) surgery on the framed link
$L=\bigcup_{i=1}^NK_i$ in $\sthree$, for an arbitrary
compact simple (gauge) Lie group $G$. Obviously, the
starting point is the partition function of CS theory
$Z(\sthree)$ on the sphere $\sthree$ [1]
$$
Z(\sthree)=\int e^{ik{\rm cs}(A)}{\cal D}A,
\eqno{\eqn{1}}
$$
where the functional integration is performed with respect
to the connections $A$ modulo gauge transformations,
defined on a trivial $G$ bundle on $\sthree$, and
$k$ is the level ($k\in\bf Z^+$). The classical action is
the {\it CS secondary characteristic class}
$$
{\rm cs}(A)={1\over4\pi}\int_\sthree\tr\left(AdA
+{2\over3}A^3\right),
\eqno{\eqn{2}}
$$
and the expectation value of
an observable $\co$ is defined as
$$
\left<\co\right>=\int\co
e^{ik{\rm cs}(A)}\cd A.
\eqno{\eqn{3}}
$$

According to the surgery prescription we should cut out a
closed tubular neighbourhood $N_i$ of $K_i$ (a solid
torus), and paste back a copy of a solid torus $T$,
matching the meridian of $T$ to the (twisted by framing
number) longitude on the boundary torus $\partial N_i$ in
$\sthree$ [5,~9]. To this end, in the first step, we should
fix boundary conditions for the field $A$ on the twisted
longitude represented by $K_i$.  Since the only
gauge-invariant (modulo conjugation) quantity defined on a
closed curve is holonomy [8], we associate the {\it
holonomy} operator $\hol{i}$ to each knot $K_i$.  Thus the
symbol
$$
\partition\eqno{\eqn{4}}
$$
should be understood as the {\it constrained} partition
function of CS theory, i.~e. the values of holonomies along
$K_i$ are fixed
$$
\hol{i}=g_i,\qquad i=1,2,\dots,N.
\eqno{\eqn{5}}
$$
Now, we can put
$$
\partition
=\left<\prodin\delta(g_i,\hol{i})\right>,
\eqno{\eqn{6}}
$$
where $\delta$ is a (group-theoretic) Dirac delta-function
[8]. Its explicit form following from the (group-theoretic)
Fourier expansion [10] is
$$
\delta(g,h)=\sigmami{}\overline{\chi_\mu(g)}\chi_\mu(h).
\eqno{\eqn{7}}
$$
Physical observables being used in CS theory are typically
Wilson loops, defined as
$$
\wilson{}=\tr_\mu(\hol{})\equiv\chi_\mu(\hol{}),
\eqno{\eqn{8}}
$$
where $\mu\in\hg$ numbers inequivalent irreducible
representations (irrep's) of $G$, and $\chi_\mu$ is a
character. By virtue of \eqn{7--8}
$$
\delta(g_i,\hol{i})=\sigmami{}\overline{\chi_\mu(g_i)}W_\mu^{K_i}(A).
\eqno{\eqn{9}}
$$
Inserting \eqn{9} into \eqn{6} yields, as a basic building
block, the following representation of the constrained
partition function
$$
\partition=\left<\prodin\sigmami{i}
\overline{\chi_{\mu_i}(g_i)}\wilson{i}\right>.
\eqno{\eqn{10}}
$$

In the second step of our construction, we should paste
back the tori matching the pairs of ``longitudes'' (the
twisted longitudes and the meridians), i.~e. we should
identify and sum up the boundary conditions. Since the
interior of a solid torus is homeomorphic to $\sthree$ with
a removed solid torus, actually the meridians play the role
of longitudes in analogous cutting procedures for an unknot
$\left\{\bigcirc\right\}$ (with reversed orientation).
Thus the partition function of CS theory on $\ml$ is
\eject
$$
Z(\ml)={1\over N_L}\int\prodin dg_i Z(\sthree,\bigcirc;g_i^{-1})
\partition\qquad\qquad\qquad
$$
$$
\qquad\qquad\qquad
={1\over N_L}\int\prodin dg_i\sigmami{i}\sum_{\nu_i\in\hg}
\overline{\chi_{\mu_i}(g_i^{-1})} \,
\overline{\chi_{\nu_i}(g_i)}
\avewilson{\mu_i}{\bigcirc}
\left<\prod_{j=1}^N W_{\nu_j}^{K_j}(A)\right>,
\eqno{\eqn{11}}
$$
where $N_L$ is a link-dependent normalization constant, and
the reversed orientation of the unknots
$\left\{\bigcirc\right\}$ (corresponding to the meridians
of the pasted back tori) accounts for the power $-1$ of the
group elements $g_i$. From the orthogonality relations for
characters and unitarity of irrep's, it follows that the
three-manifold invariant is of the form
$$
Z(\ml)={1\over
N_L}\left<\prodin\omega_{K_i}(A)\right>,
\eqno{\eqn{12a}}
$$
where
$$
\omega_{K_i}(A)\equiv\sigmami{i}\avewilson{\mu_i}{\bigcirc}
\wilson{i}
\eqno{\eqn{13}}
$$
is an element of the linear skein of an annulus, immersed
in the plane as a regular neighbourhood of $K_i$ [6]. $\langle
W_\mu^\bigcirc(A)\rangle$ are some computable coefficients depending
on $\mu$, $k$ and $G$.
Eq.~\eqn{12a} can be easily generalized to accommodate an
ordinary link ${\cal L}=\bigcup_{i=1}^M{\cal K}_i$ embedded
in $\ml$
$$
\left<\prod_{i=1}^M W_{\mu_i}^{{\cal K}_i}(A)\right>_\ml
={1\over N_L}\left<\prod_{i=1}^M W_{\mu_i}^{{\cal K}_i}(A)
\prod_{j=1}^N\omega_{K_j}(A)\right>.
\eqno{\eqn{12b}}
$$
We defer the solution of the issue of the determination of
the normalization constant $N_L$ to the end of Sect.~4.

\heading{The satellite formula}
The easiest way to calculate $\avewilson{\mu}{\bigcirc}$ follows
from the satellite formula [11,~12]. In turn, the simplest
derivation of the satellite formula on the level of skein
relations, in the context of topological field theory, could
look as follows. Let us consider the
topological-field-theory approach to skein relations, which
yields the (quasi-)braiding matrix $B$ in the form [12,~13]
$$
B=q^{\sigmaad\tensorpr{\mu}{\nu}},
\eqno{\eqn{1}}
$$
where
$$
q=e^{-{2\pi i\over k}},
\eqno{\eqn{2}}
$$
and $\mu$, $\nu$ are two irrep's of the $d$-dimensional
group $G$.  The square of
$B$, the monodromy matrix $M$ ($M=B^2$) can be derived,
for example, in the framework of the path-integral approach to
link invariants (advocated in [13]) as the contribution
coming from the intersection of the surface $\cal S$
corresponding to the representation $\mu$ and the line
$\ell$ corresponding to $\nu$. If we double the line
$\ell$, possibly assigning different representations to
each of the components, say $\nu$ and $\lambda$, there will
appear two intersection points and consequently two
contributions giving rise to
$$
B=q^{\sigmaad\tensorpr{\mu}{\nu}}\,
q^{\sigmaad\tensorpr{\mu}{\lambda}}\qquad
$$
$$
\qquad=q^{\sigmaad(\tensorpr{\mu}{\nu}\otimes\one_{\lambda}
+t_\mu^a\otimes\one_\nu\otimes t_\lambda^a)}
=q^{\sigmaad\tensorpr{\mu}{\nu\otimes\lambda}},
\eqno{\eqn{3}}
$$
where
$$
t_{\nu\otimes\lambda}^a\equiv
t_\nu^a\otimes\one_\lambda+\one_\nu\otimes
t_\lambda^a
\eqno{\eqn{4}}
$$
is a generator of $G$ in the product representation
$\nu\otimes\lambda$. Hence we have the satellite formula
(at least on the level of skein relations)
$$
W_\mu^{\cal K}(A) W_\nu^{\cal K}(A)
\approx W_{\mu\otimes\nu}^{\cal K}(A),
\eqno{\eqn{5}}
$$
where ``$\approx$'' means the ``weak equality'',
$$
X\approx Y \Longleftrightarrow
\langle X\rangle=\langle Y\rangle.
\eqno{\eqn{6}}
$$
The product on LHS of \eqn{5} should be understood in a
``regularized'' form, i.~e. the both $\cal K$'s should be
split up. Obviously, Eq.~\eqn{5} can be readily
generalized by induction to any number of factors, whereas
RHS of \eqn{5} can be expanded into irreducible components
of the product $\mu\otimes\nu$.

\heading{SU(2)-invariant}
In this section, we derive an explicit form of the
specialization of our invariant (2.12) to the group SU(2),
and show that it agrees with the result of Lickorish [6].

It appears that a very convenient way of organization of
irrep's of SU(2) group is provided by the polynomials
$S_n(x)$, closely related to the Chebyshev polynomials.
$S_n(x)$ are defined recursively by the formula
$$
S_{n+2}(x)=xS_{n+1}-S_n(x),\qquad n=0,1,\dots,
\eqno{\eqn{1a}}
$$
together with the initial conditions
$$
S_0(x)=1,\qquad S_1(x)=x.
\eqno{\eqn{1b}}
$$
Usefulness of $S_n(x)$, in the context of the SU(2) RTW
invariant, has been observed in [4,~6]. By virtue of the
definition \eqn{1}, $S_n(x)$ expresses $n$-th irrep of
SU(2) in terms of powers of the fundamental representation
$x$, denoted as $\one$ henceforth. The explicit solution of
\eqn{1} is
$$
S_n(2\cos\alpha)={\sin((n+1)\alpha)\over\sin\alpha}.
\eqno{\eqn{2}}
$$

For the group SU(2) the satellite formula (3.5) now assumes
the following elegant form
$$
W_n^{\cal K}(A)=W_{S_n(\one)}^{\cal K}(A)
\approx S_n\left(W_\one^{\cal K}(A)\right),
\eqno{\eqn{3}}
$$
whereas the skein relations for the fundamental
representation ($n=1$) of SU(2)
$$
q^{1\over4}\left<\left\{\plusskein\right\}\right>
-q^{-{1\over4}}\left<\left\{\minusskein\right\}\right>
=(q^{\half}-q^{-\half})
\left<\left\{\zeroskein\right\}\right>,
\eqno{\eqn{4a}}
$$
$$
\left<\left\{\nullskein{\pm1}\right\}\right>
=-q^{\pm{3\over4}}\left<\left\{\nullskein{0}\right\}\right>,
\eqno{\eqn{4b}}
$$
where the integers in \eqn{4b} mean a framing. Closing
the left legs of all the (three) diagrams in \eqn{4a} with
arcs, as well as the right ones, next applying \eqn{4b}, and
using the property of locality, we obtain
$$
-(q-q^{-1})\avewilson{\one}{\bigcirc}
=(q^{\half}-q^{-\half})\avewilson{\one}{\bigcirc\bigcirc}
=(q^{\half}-q^{-\half})\avewilson{\one}{\bigcirc}^2.
\eqno{\eqn{5}}
$$
Hence
$$
\avewilson{\one}{\bigcirc}=-\left(q^{\half}+q^{-\half}\right)
=-2\cos{\pi\over k},
\eqno{\eqn{6}}
$$
and by virtue of the satellite formula \eqn{3}
$$
\avewilson{n}{\bigcirc}=S_n\left(-2\cos{\pi\over k}\right)
=(-)^n{\sin{(n+1)\pi\over k}\over\sin{\pi\over k}}
=(-)^n{q^{n+1\over2}-q^{-{n+1\over2}}\over
q^{\half}-q^{-\half}}.
\eqno{\eqn{7}}
$$
We can observe a remarkable property of \eqn{7} for
$n=k-1$, namely
$$
\avewilson{k-1}{\bigcirc}=0.
\eqno{\eqn{8}}
$$
It appears that for any $\cal K$
$$
\langle\cdots W_{k-1}^{\cal K}(A)\cdots\rangle=0.
\eqno{\eqn{9}}
$$
Actually, we are dealing with a tensor algebra of finite
order, which can be interpreted as a fusion algebra [14].
In particular, Eq.~\eqn{9} immediately follows from \eqn{8}
for any $\cal K$ that can be unknotted with corresponding
skein relations.  Thus we can truncate representations
of SU(2) above the value $k-2$, and assume
$$
0\leq n\leq k-2,\qquad k=2,3,\dots.
\eqno{\eqn{10}}
$$
The final explicit form of $\omega_K$ for
the group SU(2) is then
$$
\omega_K(A)=\sum_{n=0}^{k-2}(-)^n
{q^{n+1\over2}-q^{-{n+1\over2}}\over
q^{\half}-q^{-\half}}S_n\left(W_\one^K(A)\right),
\eqno{\eqn{11}}
$$
and agrees with a corresponding expression derived by
Lickorish with a combinatorial method [6]. Strictly
speaking, $Z(\ml)$ is invariant with respect to the second
Kirby move ${\rm K}_2$. It means that it is insensitive to
the operation of sliding one of its handles over another
one. But up to now we have not considered the issue of the
determination of the normalization constant $N_L$. It
appears that proper normalization of the partition function
$Z(\ml)$ universally follows from the requirement of its
invariance with respect to the first Kirby move ${\rm
K}_1$. Hence the normalization constant $N_L$ can be chosen
in the form [6]
$$
N_L=
\left<\omega_{\bigcirc_{+1}}(A)\right>^{b_+(L)}
\left<\omega_{\bigcirc_{-1}}(A)\right>^{b_-(L)},
\eqno{\eqn{12}}
$$
where $b_+(L)$ ($b_-(L)$) is the number of positive
(negative) eigenvalues of the linking matrix of $L$.

\doublespace\noindent
{\bf Acknowledgements}

The author would like to thank Prof.~W.~B.~R.~Lickorish
for interesting discussions during his stay at the Newton
Institute, Cambridge. The author is also indebted to Prof.
H.~D.~Doebner for his kind hospitality in Clausthal. The
work was supported by the Alexander von Humboldt Foundation
and the KBN grant 202189101.

\vfill\eject

\frenchspacing
\item{1.} Witten, E., {\it Commun. Math. Phys.} {\bf 121},
351 (1989).
\item{2.} Reshetikhin, N. and Turaev, V. G., {\it Invent.
math.} {\bf 103}, 547 (1991).
\item{3.} Kirby, R. and Melvin, P., {\it Invent. math.}
{\bf 105}, 473 (1991).
\item{4.} Lickorish, W. B. R., {\it Math. Ann.} {\bf 290},
657 (1991).
\item{5.} Kauffman, L. H., {\it Knots and Physics}, World
Scientific, Singapore, 1991, Part I, Chapt. 16.
\item{6.} Lickorish, W. B. R., {\sl The skein method for
three-manifold invariants}, preprint and talk given in the
framework of the programme on {\sl Low Dimensional Topology
and Quantum Field Theory}, Cambridge, 1992.
\item{7.} Guadagnini, E., {\it Nucl. Phys. B} {\bf 375}, 381
(1992).
\item{8.} Blau, M. and Thompson, G., {\it Int. J. Mod.
Phys. A} {\bf 7}, 3781 (1992).
\item{9.} Rolfsen, D., {\it Knots and Links}, Publish or
Perish, Wilmington, 1976, Chapt. 9.
\item{10.} Wallach, N. R., {\it Harmonic analysis on homogeneous
spaces}, Marcel Dekker, 1973, Chapt. 1.
\item{11.} Morton, H. R. and Strickland, P., {\it Math. Proc.
Camb. Phil. Soc.} {\bf 109}, 83 (1991).
\item{12.} Guadagnini, E., {\it Int. J. Mod. Phys. A} {\bf
7}, 877 (1992).
\item{13.} Broda, B., {\it Mod. Phys. Lett. A} {\bf 5} 2747
(1990);
{\it Phys. Lett. B} {\bf 271}, 116 (1991).
\item{14.} Guadagnini, E., {\it Phys. Lett. B} {\bf 260},
353 (1991).

\bye